\documentclass{appolb}

\usepackage{mathptmx}

\usepackage[dvips]{color}
\usepackage{wrapfig}
\usepackage{epsf}
\usepackage{amssymb}
\usepackage{amsmath}
\usepackage{amsfonts}
\usepackage{psfrag,epsfig,graphicx,graphics}

\def\lsim{\raise0.3ex\hbox{$<$\kern-0.75em\raise-1.1ex\hbox{$\sim$}}}
\def\gsim{\raise0.3ex\hbox{$>$\kern-0.75em\raise-1.1ex\hbox{$\sim$}}}

\newcommand{\tvec}[1]{{{#1}_\perp}}

\newcommand{\rb}{\underline{r}}
\newcommand{\kb}{\underline{k}}

\newcommand{\odd}{\mathbb{O}}
\newcommand{\pom}{\mathbb{P}}

\def\bei{\begin{itemize}}
\def\ei{\end{itemize}}

\def\beeq{\begin{eqnarray}} 
\def\beqa{\begin{eqnarray}}
\def\bea{\begin{eqnarray}}

\def\eea{\end{eqnarray}}
\def\eqa{\end{eqnarray}}
\def\eeeq{\end{eqnarray}}

\def\beas{\begin{eqnarray*}}
\def\beqas{\begin{eqnarray*}}

\def\eqas{\end{eqnarray*}}
\def\eeas{\end{eqnarray*}}

\def\beq{\begin{equation}} 
\def\be{\begin{equation}}

\def\ee{\end{equation}}
\def\eq{\end{equation}}
\def\eeq{\end{equation}}

\def\beqd{\begin{displaymath}}
\def\eeqd{\end{displaymath}}
\def\eqd{\end{displaymath}}

\def\beeq{\begin{eqnarray}} \def\eeeq{\end{eqnarray}}

\newcommand{\fin}{\end{document}}

\def\bef{\begin{frame}}

\def\aut{\usebeamercolor[fg]{example text}}
\def\stmath{\usebeamercolor[fg]{structure}}

\def\stmath{}
\def\alert{}
\def\aut{}
\def\structure{}
\def\twist{}
\def\twist3{}

\def\reduced{\scriptsize}

\newcommand{\widm}{0.5\columnwidth}

\def\slashchar#1{\setbox0=\hbox{$#1$}
   \dimen0=\wd0
   \setbox1=\hbox{/} \dimen1=\wd1
   \ifdim\dimen0>\dimen1
      \rlap{\hbox to \dimen0{\hfil/\hfil}}
      #1
   \else
      \rlap{\hbox to \dimen1{\hfil$#1$\hfil}}
      /
   \fi}

\def\bei{\begin{itemize}}
\def\ei{\end{itemize}}

\def\structure#1{#1}
\def\stmath#1{#1}
\def\aut#1{#1}
\def\alert#1{#1}

\begin{document}
\title{Hard exclusive processes%
\thanks{Presented at ISMD 2012}%
}
\author{Samuel Wallon
\address{LPT, Universit{\'e} Paris-Sud, CNRS, 91405, Orsay, France {\em \&} \\
UPMC Univ. Paris 06, facult\'e de physique, 4 place Jussieu, 75252 Paris Cedex 05, France}
}
\maketitle
\begin{abstract}
We present the theory of hard exclusive processes, at medium and asymptotical energies, illustrated through some selected examples.
\end{abstract}
%\PACS{PACS numbers come here}

\section{Introduction}

\subsection{Prehistory}

Hard
exclusive processes are very efficient tools  to get insight 
into the internal tri-dimensional partonic structure of hadrons.
The idea is to reduce 
a given process to interactions involving a small number of {\it partons}
(quarks, gluons), despite confinement.
\begin{figure}[h!]
\scalebox{1.1}{\begin{tabular}{cc}
\psfrag{k}{\hspace{-.3cm}$e^-$}
\psfrag{kp}{\hspace{-.1cm}$e^-$}
\psfrag{g}{\raisebox{-.2cm}{$\hspace{-.4cm}\gamma^*$}}
\psfrag{P}{$p$}
\psfrag{Q}{\raisebox{-.2cm}{$p$}}
\psfrag{q}{$q$}
\psfrag{p}{}
\psfrag{ppq}{}
\psfrag{pro}{\scalebox{.7}{\hspace{-.5cm}\structure{hard partonic process}}}
\hspace{.4cm}\includegraphics[height=2.2cm]{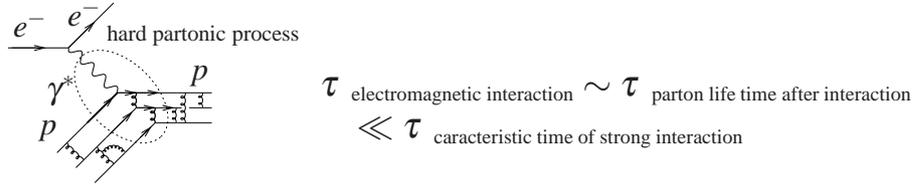} 
%\vspace{-2cm}
&
\hspace{.5cm}\raisebox{.8cm}{\scalebox{1.15}{
$\begin{array}{l}\tau_{\mbox{\, \tiny electromagnetic interaction}} \sim
\tau_{\mbox{\, \tiny parton life time after interaction}}\\
 \quad \ll \tau_{\mbox{\, \tiny caracteristic time of strong interaction}}
\end{array}$}}
\end{tabular}}
\caption{Hard subprocess for the proton form factor, with the typical time scales involved.}
\label{Fig:partonsFormFactor}
\end{figure}
This is possible if the considered process is driven by short distance phenomena, allowing the use of perturbative methods.
One should thus hit strongly enough a hadron, 
as in the case of 
an electromagnetic probe, which gives access to form factors $F_n(q^2)$
(Fig.~\ref{Fig:partonsFormFactor}).
Such exclusive reactions are very challenging since their  cross section are very small. Indeed counting rules~\cite{Brodsky:1973krBrodsky:1974vy} show that
\beq
 F_n(q^2) \simeq \frac{C}{(Q^2)^{n-1}}
\label{countingFn}
\eq
where $n$ is the number of minimal constituent (meson: $n=2$; baryons: $n=3$). 
Similarly, 
\alert{large angle} (i.e. \structure{$s \sim t \sim u$ large})  $h_a \, h_b \to h_a \, h_b $ elastic scattering satisfies~\cite{Brodsky:1981rp}, for $n$  external fermionic lines
 ($n=8 \hbox{ for }  \pi \pi \to \pi \pi$), 
% e.g. \structure{$\pi \pi \to \pi \pi$} or \structure{$p \, p \to p \, p$},
%
\beqa
\frac{d \sigma}{dt} \sim \left(\frac{\alpha_S(p_\perp^2)}s  \right)^{n-2}\,. 
\label{scaling-large-angleBL}
\eqa
\begin{figure}[h]
\centerline{\begin{tabular}{cc}
\psfrag{pq}{}
\psfrag{pg1}{}
\psfrag{pqb}{}
\psfrag{p1}{}
\psfrag{pp1}{}
\psfrag{p2}{}
\psfrag{pp2}{}
\psfrag{pu1}{}
\psfrag{pd1}{}
\psfrag{ppu1}{}
\psfrag{ppd1}{}
\psfrag{pu2}{}
\psfrag{pd2}{}
\psfrag{ppu2}{}
\psfrag{ppd2}{}
\psfrag{k}{}
\psfrag{kp}{}
\hspace{1cm}\includegraphics[height=2.5cm]{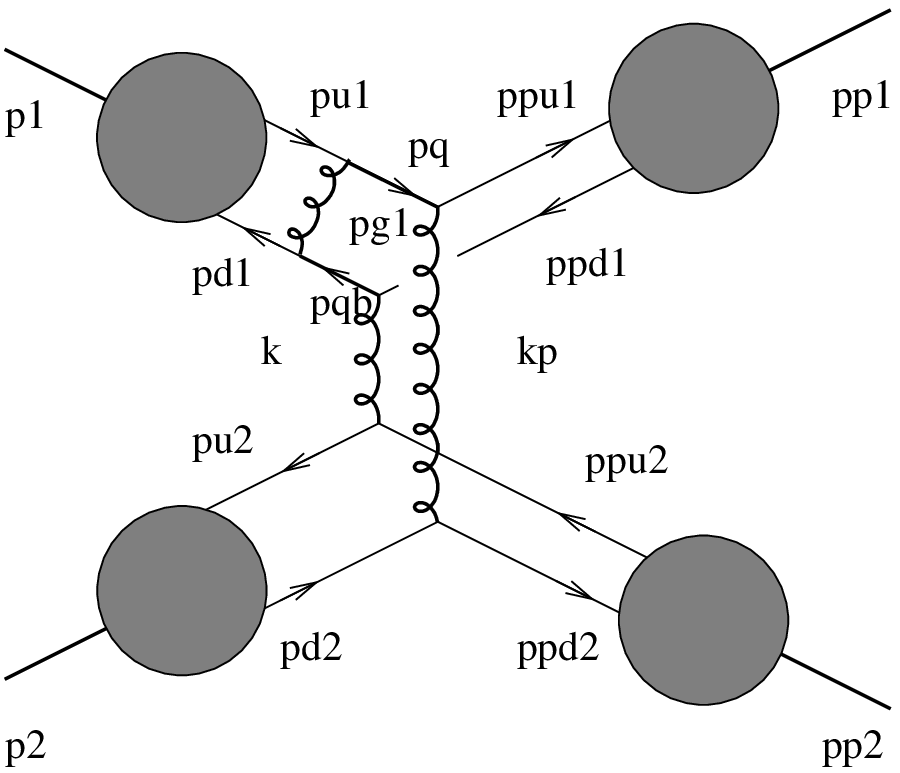} \quad \raisebox{1cm}{(a)}& 
\psfrag{pq}{}
\psfrag{pg1}{}
\psfrag{pqb}{}
\psfrag{p1}{}
\psfrag{pp1}{}
\psfrag{p2}{}
\psfrag{pp2}{}
\psfrag{pu1}{}
\psfrag{pd1}{}
\psfrag{ppu1}{}
\psfrag{ppd1}{}
\psfrag{pu2}{}
\psfrag{pd2}{}
\psfrag{ppu2}{}
\psfrag{ppd2}{}
\psfrag{k}{}
\psfrag{kp}{}
\hspace{3cm}\includegraphics[height=2.5cm]{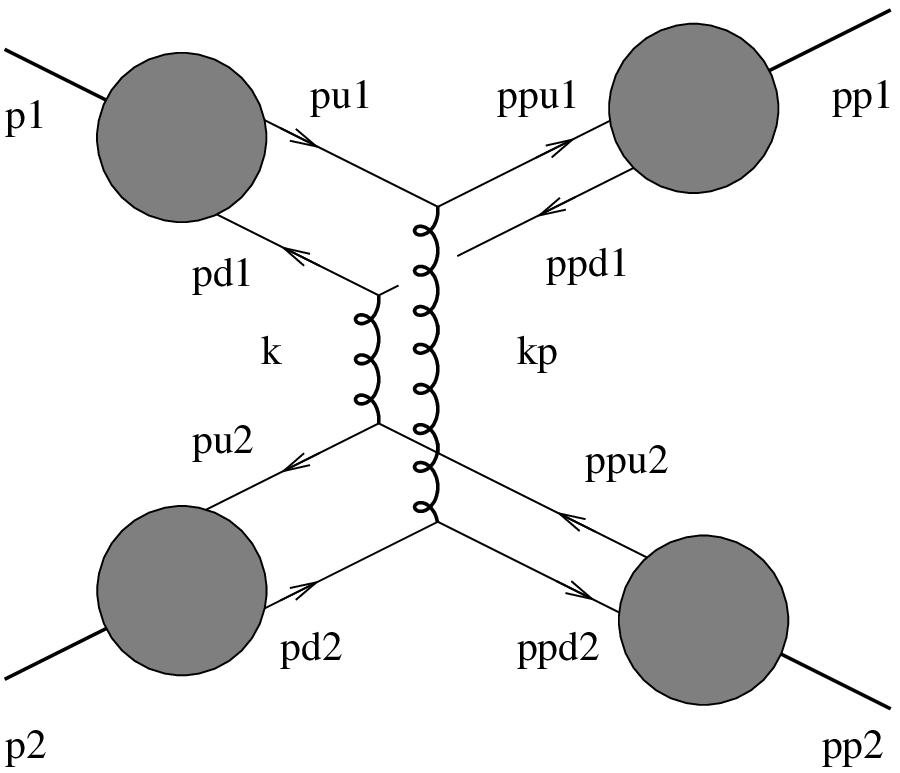} \quad \raisebox{1cm}{(b)}
\end{tabular}}
\caption{Brodsky-Lepage (a) and  Landshoff (b) mechanisms
for $\pi \pi \to \pi \pi$ at large angle.}
\label{Fig:BL-Landshoff}
\end{figure}
Limitations to the underlying factorised description have been known since decades, 
since other contributions might be significant, even at large angle~\cite{Landshoff:1974ew}. Consider~for\linebreak  example
the process \structure{$\pi \pi \to \pi \pi$}.
The  mechanism of Fig.~\ref{Fig:BL-Landshoff}a
relies on the description\linebreak  of each mesons through their collinear $q \bar{q}$ 
content, 
encoded in their distribution amplitudes (DA), the whole amplitude scaling like
$\frac{d\sigma_{BL}}{dt} \sim  s^{-6}$.
%(Eq.~(\ref{scaling-large-angleBL})). 
On~the other hand, 
one can assume\footnote{Such a mechanism is absent when at least one $\gamma^{(*)}$ is involved, due to
its point-like coupling.}
% enforcing the presence of an additional gluon as in Fig.~\ref{Fig:BL-Landshoff}a.} 
that particular collinear quark configurations of non-perturbative origin are present inside each meson (Fig.~\ref{Fig:BL-Landshoff}b),  
with a scaling
$\frac{d\sigma_{L}}{dt} \sim  s^{-5}\,.$

\subsection{Modern developments}

Inclusive and exclusive processes differ due to the hard scale power suppression, making the measurements much
more involved. This requires
high luminosity accelerators and high-performance detection facilities, as provided by
HERA (H1, ZEUS), HERMES, JLab@6 GeV (Hall A, CLAS), BaBar, Belle, BEPC-II (BES-III), LHC or by future projects (COMPASS-II, JLab@12 GeV, LHeC, EIC, ILC).
In parallel, 
theoretical efforts have been 
very important during the last decade, dealing both with perturbative and power corrections, and
popularising many new acronyms and concepts
which we now introduce in a nutshell\footnote{For reviews, see~\cite{Guichon:1998xvGoeke:2001tz, Diehl:2003ny, Belitsky:2005qn, Boffi:2007ycBurkert:2007zzGuidal:2008zzaWallon:2011zx}.}.

%%%%%%%%%%%%%%%%%%%%%%%%%%%%%%%%%%%%%%%%%%%%%%%

%%%%%%%%%%%%%%%%%%%%%%%%%%%%%%%%%%%%%%%%%%%%%%%

\section{Collinear factorisations}

\subsection{Extensions from DIS}

\begin{figure}[h!]
\scalebox{.88}{\centerline{\hspace{1.2cm}\begin{tabular}{cc}
\raisebox{-.44 \totalheight}
{\psfrag{ph1}{$\gamma^*$}
\psfrag{ph2}{$\gamma^*$}
\psfrag{s}{$s$}
\psfrag{t}{$t$}
\psfrag{hi}{$p$}
\psfrag{hf}{$p$}
\psfrag{Q1}{\raisebox{-.2cm}{\,\,$Q^2$}}
\psfrag{Q2}{\raisebox{-.2cm}{\hspace{-.3cm}$Q^2$}}
\psfrag{x1}{$x$}
\psfrag{x2}{$x$}
\psfrag{GPD}{\raisebox{-.05cm}{\hspace{-.1cm}PDF}}
\psfrag{CF}{\raisebox{-.05cm}{\hspace{-.1cm}CF}}
\hspace{0cm}\scalebox{.8}{\raisebox{0 \totalheight}{\hspace{2cm}\includegraphics[height=5cm]{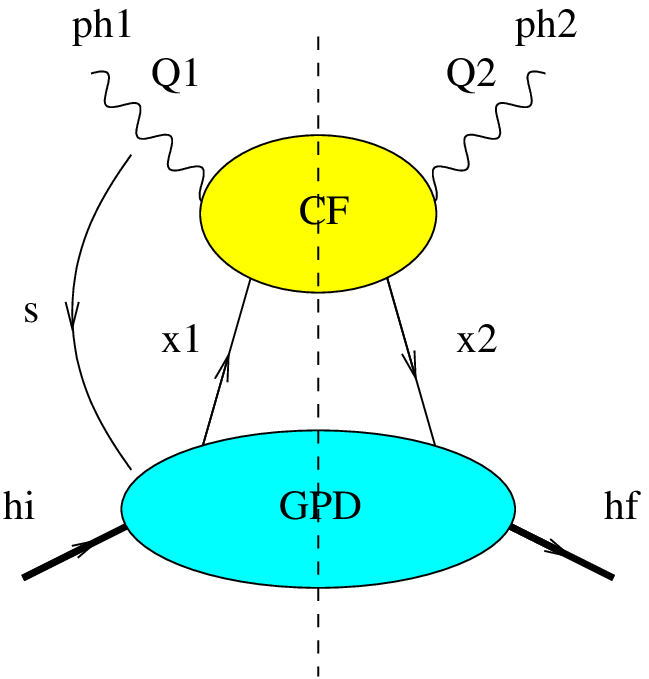}}}}
\raisebox{-.44 \totalheight}
{\psfrag{ph1}{$\gamma^* \, [\gamma]$}
\psfrag{ph2}{$\gamma\, [\gamma^*]$}
\psfrag{s}{$s$}
\psfrag{t}{$t$}
\psfrag{hi}{$p$}
\psfrag{hf}{$p'$}
\psfrag{Q1}{\raisebox{-.2cm}{\,\,$Q^2$}}
\psfrag{Q2}{\raisebox{-.2cm}{\hspace{-.5cm}$[Q^2]$}}
\psfrag{GPD}{\raisebox{-.05cm}{\hspace{-.1cm}GPD}}
\psfrag{CF}{\raisebox{-.05cm}{\hspace{-.1cm}CF}}
\psfrag{x1}{\hspace{-.4cm}$x+\xi$}
\psfrag{x2}{$x-\xi$}
\hspace{2.5cm}\scalebox{.8}{\raisebox{0 \totalheight}{\includegraphics[height=5cm]{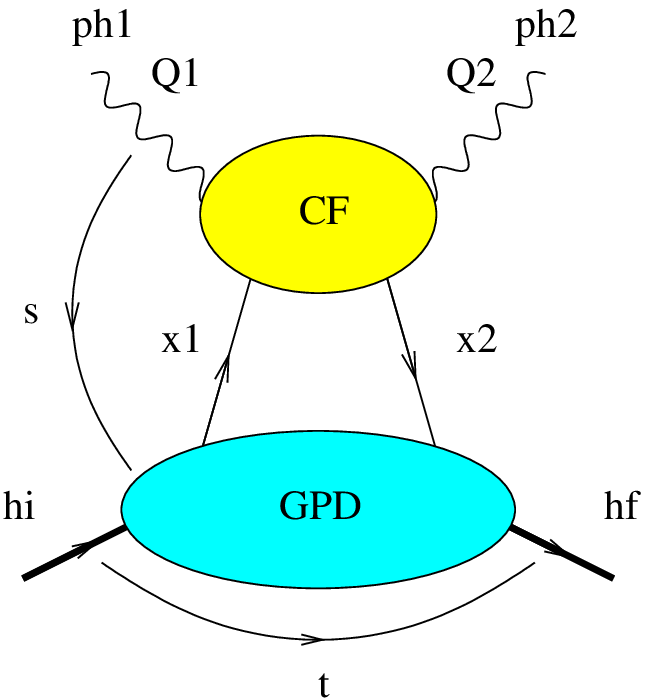}}}}
%{\includegraphics[height=2.8cm]{dvcs-eng.ps}} 
\end{tabular}}}
% \vspace{-.5cm}
% 
\caption{(a): DIS factorisation.  (b): DVCS [TCS] factorisation.}
\label{Fig:factDIS-DVCS}
\end{figure}

 The deep inelastic scattering (DIS) total cross-section, as an {\it inclusive} process,  involves the \structure{forward} ($t=0$) Compton amplitude, through optical theorem (Fig.~\ref{Fig:factDIS-DVCS}a).
 The structure functions can be factorised collinearly as a convolution of coefficient functions (CFs) with parton distribution functions (PDFs).
The {\it exclusive} deep virtual Compton scattering (DVCS)  
and time-like Compton scattering (TCS),
 in the limit 
$s_{\gamma^* p}, \, Q^2 \gg -t\,,$ can also be factorised,  
now at the amplitude level (Fig.~\ref{Fig:factDIS-DVCS}b). It involves generalised parton distribution functions (GPDs) 
\cite{Mueller:1998fvRadyushkin:1996ndJi:1996ek} which extend the PDFs outside of the diagonal kinematical limit: the $t$ variable as well as the longitudinal momentum transfer may not vanish, calling for new variables, the skewness $\xi$,
encoding the inbalance of longitudinal
$t-$channel momentum, and the 
transfered transverse momentum $\Delta$.
\psfrag{g}{$\gamma$}
\psfrag{gs}{$\gamma^*$}
\psfrag{s}{$s$}
\psfrag{t}{$t$}
\psfrag{Q2}{$\!\!\!Q^2$}
\psfrag{CF}{\!\!\reduced  CF}
\begin{figure}[h!]
\centerline{\scalebox{.77}{\centerline{\begin{tabular}{cc}
\psfrag{ph1}{$\gamma^*$}
\psfrag{s}{$s$}
\psfrag{t}{$t$}
\psfrag{hi}{$h$}
\psfrag{hf}{$h'$}
\psfrag{Q1}{$Q^2$}
\psfrag{GPD}{\raisebox{-.05cm}{\hspace{-.1cm}GPD}}
\psfrag{CF}{\raisebox{-.05cm}{\hspace{-.1cm}CF}}
\psfrag{DA}{\raisebox{.15cm}{\rotatebox{-55}{DA}}}
\psfrag{x1}{\hspace{-.4cm}$x+\xi$}
\psfrag{x2}{$x-\xi$}
\psfrag{z}{$u$}
\psfrag{zb}{$-\bar{u}$}
\psfrag{rho}{$\rho, \, \pi$}
\hspace{-.5cm}
\raisebox{-.44 \totalheight}{\includegraphics[height=4.8cm]{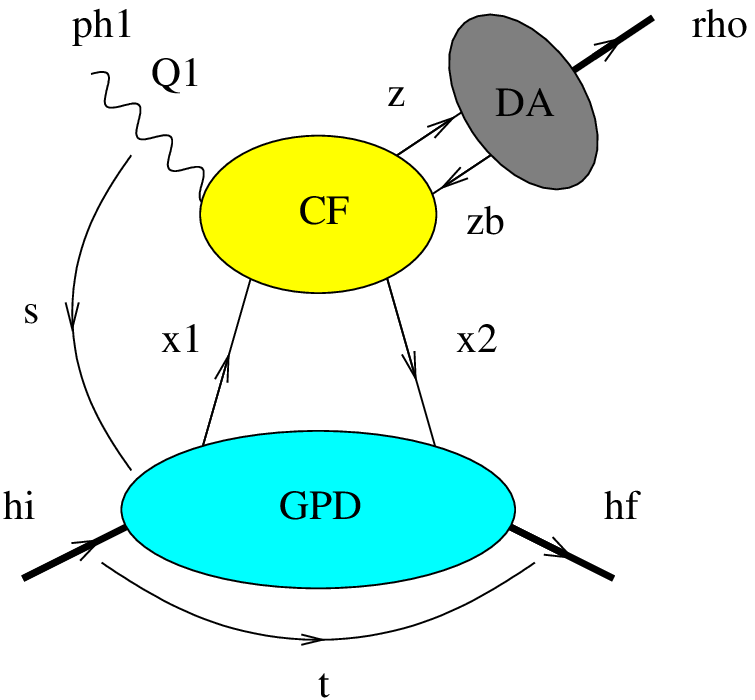}}
&
\psfrag{h}{ hadron}
\psfrag{GDA}{\raisebox{-.05cm}{\hspace{-.12cm}GDA}}
\psfrag{CF}{\raisebox{-.05cm}{\hspace{-.1cm}CF}}
\hspace{1cm}\raisebox{-.4 \totalheight}{\includegraphics[height=4.5cm]{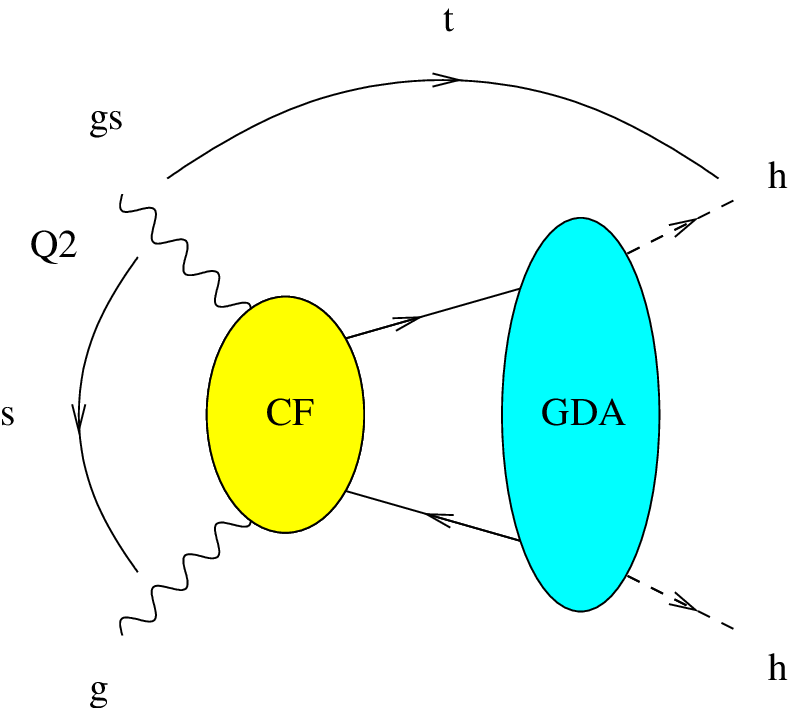}}
\end{tabular}}}}
\caption{(a): Collinear factorisation of meson electroproduction.
 (b): Collinear factorisation of hadron pair production in $\gamma \gamma^*$ subchannel.}
\label{Fig:factRho-double_meson}
\end{figure}

\begin{figure}
\centerline{
\scalebox{.8}{\begin{tabular}{ccc}
\psfrag{ph1}{$\gamma^*$}
\psfrag{ph2}{$\gamma$}
\psfrag{s}{$s$}
\psfrag{t}{$t$}
\psfrag{hi}{$h$}
\psfrag{hf}{$h'$}
\psfrag{Q1}{$Q^2$}
\psfrag{Q2}{}
\psfrag{GPD}{\raisebox{-.05cm}{\hspace{-.1cm}GPD}}
\psfrag{CF}{\raisebox{-.05cm}{\hspace{-.1cm}CF}}
\psfrag{x1}{\hspace{-.4cm}$x+\xi$}
\psfrag{x2}{$x-\xi$}
\raisebox{-.44 \totalheight}{\includegraphics[height=5cm]{dvcs.eps}}
&
\scalebox{1.3}{$\stackrel{t \, \to \, u}{\longrightarrow}$}
&
\psfrag{hi}{$h$}
\psfrag{hf}{$h'$}
\psfrag{Q1}{$Q^2$}
\psfrag{Q2}{}
\psfrag{s}{$s$}
\psfrag{t}{$t$}
\psfrag{ph1}{$\gamma^*$}
\psfrag{ph2}{$\gamma$}
\psfrag{x1}{\hspace{-.4cm}$x+\xi$}
\psfrag{x2}{$x-\xi$}
\psfrag{GPD}{\raisebox{-.05cm}{\hspace{-.1cm}TDA}}
\psfrag{CF}{\raisebox{-.05cm}{\hspace{-.1cm}CF}}
\raisebox{-.44 \totalheight}{\includegraphics[height=5cm]{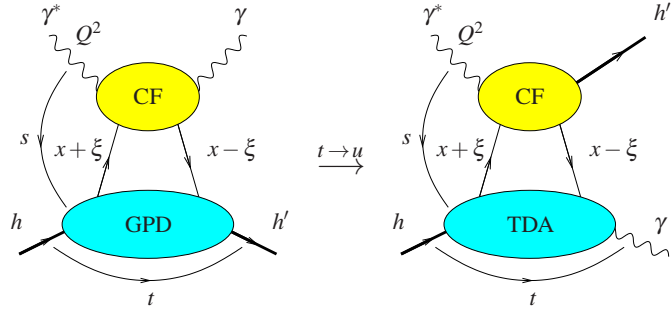}}
\end{tabular}}}
\caption{$t \leftrightarrow u$ crossing from DVCS.
}
\label{Fig:factTDA}
\end{figure}

From DVCS, several extensions have been made. First, one may replace the produced $\gamma$ by a meson, factorised collinearly through a DA~\cite{Collins:1996fbRadyushkin:1996ru} (Fig.~\ref{Fig:factRho-double_meson}a).  Second, one may consider the crossed process 
in the limit $s_{\gamma^* p}, \, Q^2 \ll -t$. It again factorises (Fig.~\ref{Fig:factRho-double_meson}b), the $q \, \bar{q}$ content of the hadron pair being encoded in a genera-lised distribution amplitude (GDA)~\cite{Diehl:1998dk}. These frameworks allow to describe 
hard exotic hybrid meson production both in electroproduction  and $\gamma \gamma^*$ collisions~\cite{Anikin:2004vcAnikin:2004jaAnikin:2006du}.
Starting from usual DVCS, one can allow the initial hadron and the final hadron to differ, 
replacing GPDs by transition GPDs. 
The conservation of baryonic number can be removed between inital and final state, introducing transition distribution amplitudes (TDAs)~\cite{Pire:2004ie}. This can be obtained from 
DVCS by a $t \leftrightarrow u$ crossing (Fig.~\ref{Fig:factTDA}). 
%
%%%
\psfrag{p}{$p$}  
\psfrag{pp}{$p'$}
\psfrag{q}{$q$}
\psfrag{qp}{$q'$}
\psfrag{pip}{$p$}
\psfrag{pim}{$\pi^-$}
\psfrag{pr}{$p$}
\psfrag{apr}{$\bar p$}
\psfrag{g}{$\gamma$}
\psfrag{gs}{$\gamma^*$}
\psfrag{u}{$u$}
\psfrag{db}{$\bar d$}
\psfrag{ep}{$e^+$}
\psfrag{em}{$e^-$}
\psfrag{d}{$d$} 
\psfrag{a}{$a$}
\psfrag{b}{$b$}
\psfrag{pi}{$\pi$}
\psfrag{k}{$k$}
\psfrag{DA}{$\,\,DA$}
\psfrag{TDA}{$TDA$}
\psfrag{TH}{$T_H$}
\psfrag{q}{}
\psfrag{pim}{}
A further extension is done by replacing the outoing \alert{$\gamma$} by any hadronic state~\cite{Pire:2005axLansberg:2007se}. 
The process $\gamma^* \, \gamma \to \rho \, \rho$ is of particular interest, since it  can be factorised in two ways involving either the GDA of the $\rho$ pair or the $\gamma^* \to \rho$
TDA, depending on the polarization of the incoming photons~\cite{Pire:2006ik}.

%%%%%%%%%%%%%%%%%%%%%%%%%%%%%%%%%%%%%%%%%%%%%%%%%%%%%%%%%%%%%%%%%%%%

\subsection{GPDs}

\begin{figure}[h!]
\begin{center}
     \leavevmode
     \epsfxsize=.91\textwidth

     \epsffile{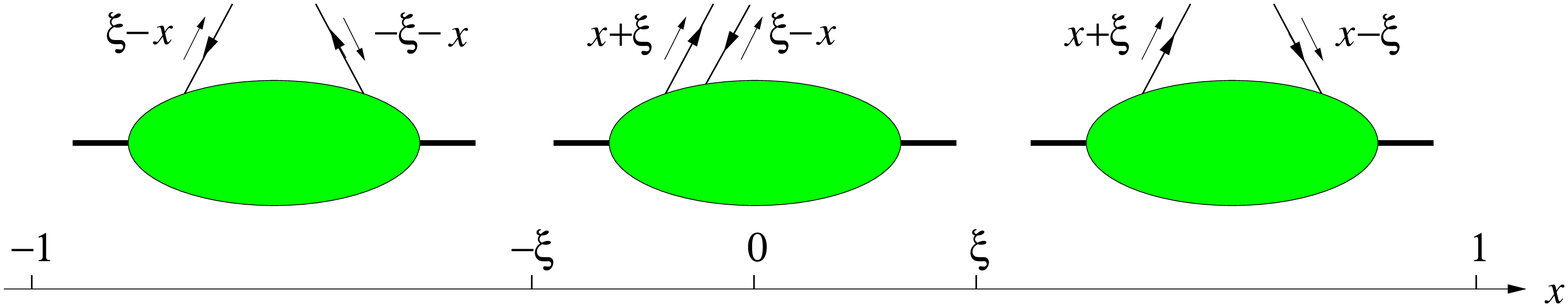}
\end{center}
\begin{tabular}{ccc}
\hspace{0.5cm}
\begin{tabular}{c}
\structure{\footnotesize Emission and reabsoption} \\
\structure{\footnotesize of an antiquark}\\
\structure{\footnotesize $\sim$ PDFs for antiquarks}\\
\quad {\aut DGLAP}-II region
\end{tabular}
&
\hspace{-.6cm}
\begin{tabular}{c}
\structure{\footnotesize Emission of a quark and} \\
\structure\footnotesize {emission of an antiquark}\\
\structure{\footnotesize $\sim$ meson exchange}\\
\quad {\aut ERBL} region
\end{tabular}
&
\hspace{-.6cm}
\begin{tabular}{c}
\structure{\footnotesize Emission and reabsoption} \\
\structure{\footnotesize of a quark}\\
\structure{\footnotesize $\sim$ PDFs for quarks}\\
\quad {\aut DGLAP}-I region
\end{tabular}
\end{tabular}
\caption{\label{Fig:regions_GPD} The parton interpretation of GPDs in
 the three $x$-intervals. Figure from~\cite{Diehl:2003ny}.
}
\end{figure}

The twist 2 GPDs have a simple
physical interpretation, shown in Fig.~\ref{Fig:regions_GPD}. Their
classification goes as follows, according to the chirality of the $\Gamma$ matrix involved in the matrix elements $F^q$ and $\tilde{F}^q$ of bilocal light-cone operators defining them:

\bei

\item For quarks, one should distinguish the exchanges
	\bei
	\item \structure{without helicity flip} (\structure{chiral-even} \alert{$\Gamma$} matrices), \alert{4 chiral-even GPDs} :

\alert{$H^q$} ($\xrightarrow{\tiny\xi=0,\, t=0}$ \structure{PDF} $q$) \alert{, $E^q$, $\tilde{H}^q$} $(\xrightarrow{\tiny\xi=0,\,t=0}$ \structure{polarized PDFs} $\Delta q$) and \alert{$\tilde{E}^q$},
%%%%%%%
\beqas
  \label{GPD_quark}
&&\hspace{-2.2cm} F^q =
\frac{1}{2} \int \frac{d z^+}{2\pi}\, e^{ix P^- z^+}
  \langle p'|\, \bar{q}(-\half z)\, {\alert{\gamma^-}} q(\half z) 
  \,|p \rangle \Big|_{z^-=0,\,\, \tvec{z}=0}
\nonumber \\
&&\hspace{-2.2cm} \phantom{F^q} = \frac{1}{2P^-} \left[
  \alert{H^q}(x,\xi,t)\, \bar{u}(p') \gamma^- u(p) +
  \alert{E^q}(x,\xi,t)\, \bar{u}(p') 
                 \frac{i \,\sigma^{-\alpha} \Delta_\alpha}{2m} u(p)
  \, \right] ,
\nonumber \\
&&\hspace{-2.2cm}\tilde{F}^q =
\frac{1}{2} \int \frac{d z^+}{2\pi}\, e^{ix P^- z^+}
  \langle p'|\, 
     \bar{q}(-\half z)\, \alert{\gamma^- \gamma_5}\, q(\half z)
  \,|p \rangle \Big|_{z^-=0,\, \,\tvec{z}=0}
\nonumber \\
&&\hspace{-2.2cm} \phantom{F^q} = \frac{1}{2P^-} \left[
 \alert{\tilde{H}^q}(x,\xi,t)\, \bar{u}(p') \gamma^- \gamma_5 u(p) +
  \alert{\tilde{E}^q}(x,\xi,t)\, \bar{u}(p') \frac{\gamma_5 \,\Delta^-}{2m} u(p)
  \, \right] .
\eqas

	\item
	\structure{with helicity flip} ( \structure{chiral-odd} \alert{$\Gamma$}  mat.), 
\alert{4 chiral-odd GPDs:} 

$H^q_T$ ($\xrightarrow{\tiny\xi=0,\,t=0}$ \structure{quark transversity PDFs} $\Delta_T q$\alert{), $E^q_T$, $\tilde{H}^q_T$, $\tilde{E}^q_T$}
\beqas
&&\hspace{-2.2cm}\frac{1}{2} \int \frac{d z^+}{2\pi}\, e^{ix P^- z^+}
  \langle p'|\, 
     \bar{q}(-\half z)\, i \, \alert{\sigma^{-i}}\, q(\half z)\, 
  \,|p \rangle \Big|_{z^-=0,\,\, \tvec{z}=0} 
 \\
&&\hspace{-2.21cm}= \!\frac{1}{2P^-} \bar{u}(p') \! \!\!\left[\!
 \alert{H_T^q}\, i \sigma^{-i} \!\!\!+\!
  \alert{\tilde{H}_T^q}\, \frac{P^- \Delta^i - \Delta^- P^i}{m^2} \!+\!
  \alert{E_T^q}\, \frac{\gamma^- \Delta^i - \Delta^- \gamma^i}{2m} \!+\!
  \alert{\tilde{E}_T^q} \frac{\gamma^- P^i - P^- \gamma^i}{m}
  \right] \!\! u(p)  , \nonumber
\eqas
	\ei

\item A similar analysis can be made for twist-2 gluonic GPDs:

	\bei

	\item \alert{4 gluonic GPDs} \structure{without helicity flip:} 

\alert{$H^g$} ($\xrightarrow{\tiny\xi=0,\,t=0}$ \structure{PDF} $x \, g$), 
\alert{$E^g$},
\alert{$\tilde{H}^g$} ($\xrightarrow{\tiny\xi=0,\,t=0}$ \structure{polarized PDF} $x \, \Delta g$) and
\alert{$\tilde{E}^g$}

	\item \alert{4 gluonic GPDs} \structure{with helicity flip:} 
\alert{$H^g_T$},
\alert{$E^g_T$},
\alert{$\tilde{H}^g_T$} and 
\alert{$\tilde{E}^g_T$}.
We note that there is 
no forward limit reducing to gluons PDFs here: a change of  2 units of helicity cannot be  compensated by a spin 1/2 target.
	\ei

\ei

%%%%%%%%%%%%%%%%%%%%%%%%%%%%%%%%%%%%%%%%%%%%%%%%%%%%%%%%
%%%%%%%%%%%%%%%%%%%%%%%%%%%%%%%%%%%%%%%%%%%%%%%%%%%%%%%%

%
\subsection{Transversity}

The tranverse spin content of the proton is related to non-diagonal helicity observables, since
\beqas
%\centerline{
\hbox{spin along }  x : \quad
\begin{tabular}{ccc}$| \uparrow \rangle_{(x)}  $  & $\sim$ & $| \rightarrow\rangle +| \leftarrow\rangle$\\
$| \downarrow \rangle_{(x)}  $ & $\sim$ & $| \rightarrow\rangle -| \leftarrow\rangle$ 
\end{tabular} \quad \hbox{: helicity states}
\,.
\eqas
An observable sensitive to helicity spin flip
gives thus access to the transversity
\alert{$\Delta_T q(x)$},
which is very badly known.
Meanwhile, the transversity GPDs are completely unknown.
Since for massless (anti)quarks chirality = (-) helicity,  
transversity is a chiral-odd quantity.
Now, since QCD and QED are chiral even, any chiral-odd operator should be balanced by another chiral-odd operator in the amplitude.
\begin{figure}[h!]
\psfrag{z}{\hspace{-0.1cm}\footnotesize $z$ }
\psfrag{zb}{\raisebox{0cm}{\hspace{-0.1cm} \footnotesize$\bar{z}$} }
\psfrag{gamma}{\raisebox{+.1cm}{\footnotesize $\,\gamma$} }
\psfrag{pi}{\footnotesize$\!\pi$}
\psfrag{rho}{\footnotesize$\,\rho$}
\psfrag{TH}{\raisebox{-.05cm}{\hspace{-0.1cm}\footnotesize $T_H$}}
\psfrag{tp}{\raisebox{.3cm}{\footnotesize $t'$}}
\psfrag{s}{\hspace{0.4cm}\footnotesize$s$ }
\psfrag{Phi}{}%{ \hspace{-0.3cm} $\phi$}
%\hspace{-2cm}
\centerline{\hspace{-2.7cm} 
\scalebox{.9}{\raisebox{.6cm}{\includegraphics[width=4.3cm]{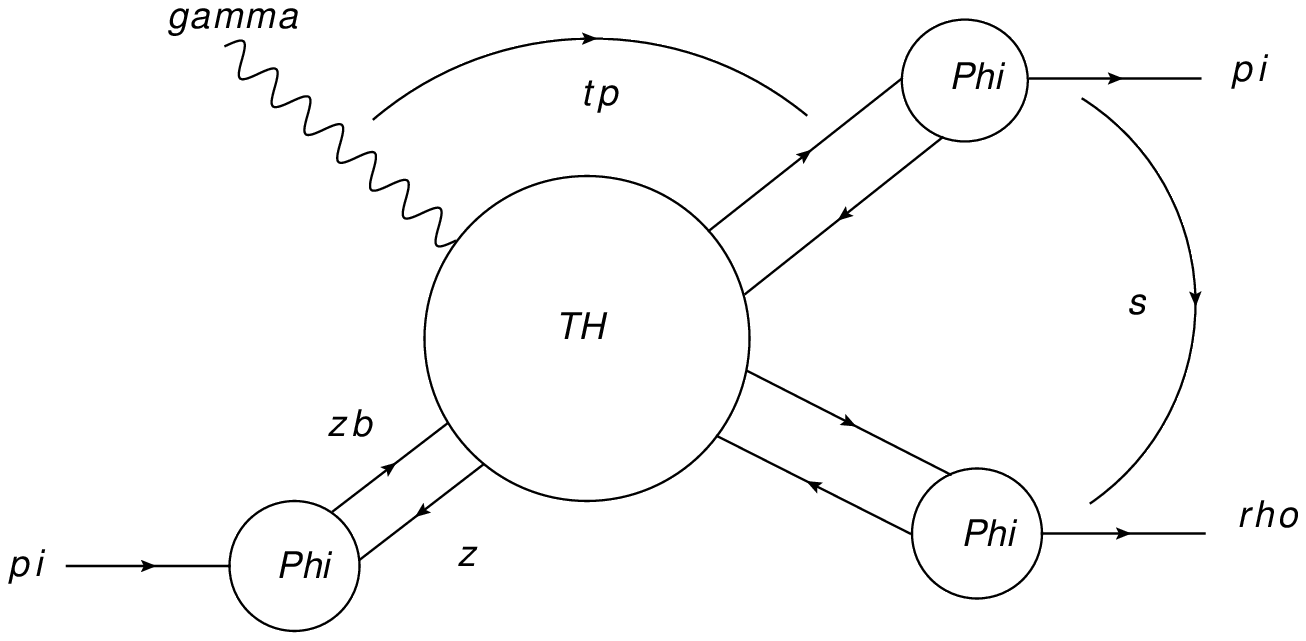}}}\hspace{0.5cm} \raisebox{1.5cm}{$\longrightarrow$}\hspace{0.2cm}
\psfrag{piplus}{\footnotesize$\,\pi^+ $ chiral-\alert{even} twist 2 DA}
\psfrag{rhoT}{\footnotesize$\,\rho^0_T$ chiral-\alert{odd} twist 2 DA}
\psfrag{M}{\hspace{-0.15cm} \footnotesize $M^2_{\pi \rho}$ }
\psfrag{x1}{\raisebox{-.1cm}{\hspace{-0.6cm} \footnotesize $x+\xi $  }}
\psfrag{x2}{\raisebox{-.1cm}{\hspace{-0.1cm} \footnotesize  $x-\xi $ }}
\psfrag{N}{ \hspace{-0.4cm}\footnotesize $N$}
\psfrag{GPD}{\raisebox{-.1cm}{\footnotesize \hspace{-0.4cm}  $GPDs$}}
\psfrag{Np}{\footnotesize$N'$}
\psfrag{t}{ \raisebox{-.1cm}{\footnotesize \hspace{-0.3cm}   $t \ll M^2_{\pi \rho}$  \hspace{.1cm}\raisebox{-.1cm}{chiral-\alert{odd} twist 2 GPD}}}
\scalebox{.9}{\includegraphics[width=4.5cm]{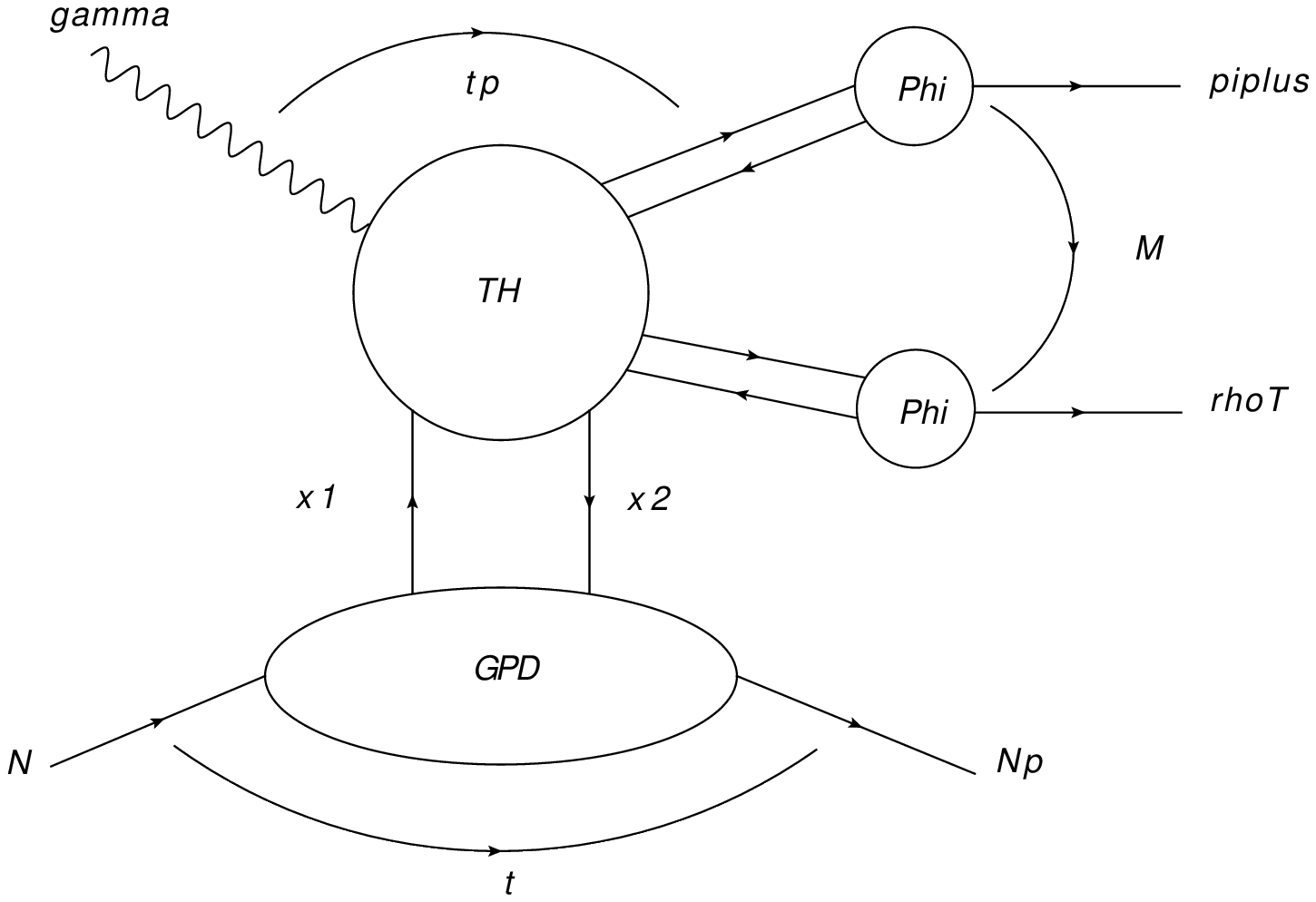}}}
\vspace{.2cm}
\caption{Brodsky-Lepage factorisation applied to $\gamma \, N \to \pi^+ \, \rho^0_T \, N'$.}
\label{Fig:transversity3}
\end{figure}
The dominant DA for $\rho_T$ is of twist 2 and chiral-odd. One may thus think about using $\rho_T$-electroproduction.
Unfortunately
the amplitude vanishes,
at any order in perturbation theory, since this process would require a transfer of 2 units of helicity from the proton~\cite{Diehl:1998pdCollins:1999un}.
This vanishing is true only a twist 2,
however processes involving twist 3 DAs~\cite{Ball:1998skBall:1998ffBall:2007zt} may face problems with factorisation
(see Sec.~\ref{SubSec:PB}).
One can circumvent this vanishing by considering a 3-body final state~\cite{Ivanov:2002jjEnberg:2006heBeiyad:2010cxa}.
Indeed the process
$\gamma \, N \to \pi^+ \, \rho^0_T \, N'$ can be described in the spirit of large angle factorisation~\cite{Brodsky:1981rp} of the process
$\gamma \, \pi \rightarrow \pi \, \rho $ \structure{at large $s$ and fixed angle} (i.e. for fixed $t'/s, \, u'/s$ in Fig.~\ref{Fig:transversity3}), $M_{\pi\rho}^2$ providing the hard scale.
Such processes with a 3-body final state can give access to all GPDs,
$M_{\pi\rho}^2$ playing the role of the $\gamma^*$ virtuality of usual TCS.

%%%%%%%%%%%%%%%%%%%%%%%%%%%%%%%%%%%%DVCS, TCS, resummation 

%%%%%%%%%%%%%%%%%%%%%%%%%%%%%%%%%%%%%%%%%%%%%%

\subsection{Resummation effects}

%%%%%%%%%%%%%%%%%%%%%%%%%%%%%%%%%%%%%%%%%%%%%%%%%%%%%%%%%%%%%%%%%%%%%%%%
The DVCS coefficient function has  threshold singularities in its $s-$ and $u$-channels, in the limits $x \to \pm \xi\,.$ Soft-collinear effects lead to large terms of type
$[\alpha_S \log^2(\xi\pm x)]^n/(x\pm \xi)$ which can be resummed in light-like gauge as ladder-like diagrams~\cite{Altinoluk:2012ntAltinoluk:2012fb}.

\subsection{Problems with factorisation}
\label{SubSec:PB}

In $\rho-$electroproduction, since QED and QCD vertices are chiral even,
the total helicity of a $q \bar{q}$ pair produced by a $\gamma^*$ should be 0, and the  $\gamma^*$ helicity equals $L^{q \bar{q}}_z$. In the pure collinear limit (i.e. twist 2), $L^{q \bar{q}}_z$=0, and thus the $\gamma^*$ is longitudinally polarised.
At $t=0$ there is no source of orbital momentum from the proton coupling 
so that the meson and photon helicities are identical. This statement is not modified in
the collinear factorisation approach at $t \neq 0$ (the hard part is $t-$independent). This $s-$channel helicity conservation (SCHC) implies that the only allowed transitions are 
 $\gamma^*_L \to \rho_L$, for which QCD factorisation  holds at twist 2 at any order in perturbation~\cite{Collins:1996fbRadyushkin:1996ru}, and 
$\gamma^*_T \to \rho_T$, for which QCD factorisation faces
problems due to end-point singularities at twist 3 when integrating over quark longitudinal momenta~\cite{Mankiewicz:1999tt}.
The improved collinear approximation may be a solution: one
 keeps a transverse $\ell_\perp$ dependency in the $q,$ $\bar{q}$ momenta,  to regulate end-point singularities.
 Now,
 soft and collinear gluon exchange between the valence quark are responsible for large double-logarithmic effects which are conjectured to exponentiate in a Sudakov factor~\cite{Li:1992nu}, regularizing
end-point singularities. This tail can be combined with an ad-hoc non-perturbative gaussian ansatz for the DAs, providing 
 practical tools for meson electroproduction phenomenology~\cite{Goloskokov:2005sdGoloskokov:2006hrGoloskokov:2007nt}.

%%%%%%%%%%%%%%%%%%%%%%%%%%%%%%%%%%%%

%%%%%%%%%%%%%%%%%%%%%%%%%%%%%%%%%%%%%%%%%%%%%%%%%%%%%%%%%%%%%%%%%%%%%
%%%%%%%%%%%%%%%%%%%%%%%%%%%%%%%%%%%%%%%%%%%%%%%%%%%%%%%%%%%%%%%%%%%%%

\section{QCD at large $s$}

\subsection{Theorical motivations}

The perturbative {\aut Regge} limit of QCD 
is reached 
in the diffusion of two hadrons $h_1$ and $h_2$ whenever
$\alert{\sqrt{s_{h_1 \, h_2}}}
 \gg$  other scales (masses, transfered momenta, ...), while 
other scales are comparable (virtualities, etc...) and at least one of them is large enough to justify the applicability of perturbative QCD.
The appearance of large $\ln s$ in loop corrections may compensate the smallness of $\alpha_s$. 
The dominant sub-series $\sum_n (\alpha_s \, \ln s)$
leads to
$\alert{\sigma_{tot}^{h_1\, h_2}}  \sim \alert{ s^{\alpha_\pom(0) -1}}\,,$
($\alpha_\pom(0) >1$)~\cite{Fadin:1975cbKuraev:1976geKuraev:1977fsBalitsky:1978ic}
which violates  QCD $S$ matrix \alert{unitarity}. One of the
main issue of QCD is to improve this result, and to test this dynamics experimentally, now in particular based on exclusive processes.

%%%%%%%%%%%%%%%%%%%%%%%%%%%%%%%%%

\subsection{$k_T$-factorisation}

\begin{figure}
\psfrag{g1}[cc][cc]{$\gamma^*(q_1)$}
\psfrag{g2}[cc][cc]{$\gamma^*(q_2)$}
\psfrag{p1}[cc][cc]{\raisebox{.3cm}{$\,\rho(p_1)$}}
\psfrag{p2}[cc][cc]{\raisebox{.3cm}{$\,\rho(p_2)$}}

\psfrag{l1}[cc][cc]{}%{$l_1$}
\psfrag{l1p}[cc][cc]{}%{$-\tilde{l}_1$}
\psfrag{l2}[cc][cc]{}%{$l_2$}
\psfrag{l2p}[cc][cc]{}%{$-\tilde{l}_2$}
\psfrag{ai}[cc][cc]{$\stmath \beta^{\,\nearrow}$}
\psfrag{bd}[cc][cc]{$\stmath \alpha_{\,\searrow}$}
\psfrag{k}[cc][cc]{$k$}
\psfrag{rmk}[lc][cc]{$\hspace{-.5cm}r-k \hspace{3cm} \int d^2 k_\perp$}
%\psfrag{oa}[cc][cc]{\raisebox{0.46
   % \totalheight}{${\tiny \alpha_k \ll \alpha_{\rm quarks}}$}}
\psfrag{oa}[cc][cc]{\scalebox{.9}{${}\quad {\tiny \stmath \alpha \ll \alpha_{\rm quarks}}$}}
\psfrag{ou}[ll][ll]{%$\slashchar{p}_1$ \quad 
\hspace{1cm} $\Rightarrow$ set $\alpha=0$ and  $\int d\beta$ }
\psfrag{ob}[cc][cc]{\scalebox{.9}{\raisebox{-1.5
    \totalheight}{$\quad {\tiny \stmath\beta \ll \beta_{\rm quarks}}$}}}
 %\psfrag{ob}[cc][cc]{${} \quad {\tiny \stmath\beta_k \ll \beta_{\rm quarks}}$}
\psfrag{od}[ll][ll]{%$\slashchar{p}_2$ \quad 
\hspace{1cm} $\Rightarrow$ set $\beta=0$ and  $\int d\alpha$}
\scalebox{1}{
\centerline{\hspace{-3cm}\epsfig{file=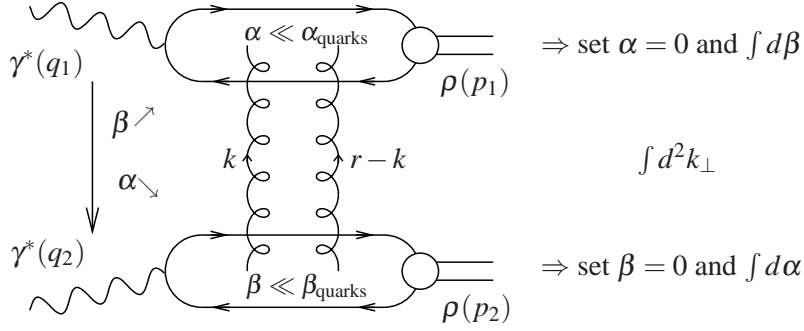,width=\widm}}
}
\caption{$k_T-$factorisation applied to $\gamma^* \, \gamma^* \to \rho \, \rho\,.$}
\label{Fig:kT-factorization}
\end{figure}

The main tool in this regime is the $k_T$-factorisation, as illustrated in Fig.~\ref{Fig:kT-factorization} for $\gamma^* \, \gamma^* \to \rho \, \rho$.
Using the {\aut Sudakov} decomposition
$k = \alpha \,p_1 + \beta \, p_2 + k_\perp$ (with $p_1^2=p_2^2=0,\,  2 p_1 \cdot p_2=s$), in which
${\stmath d^4k= \frac{s}{2} \, d \alpha \, d\beta \, d^2k_\perp}\,,$ and noting that the dominant 
polarization of the $t-$channel gluons is \alert{non-sense}, i.e.
$\varepsilon_{\alert{NS}}^{up}=\frac{2}s \, p_2$, $\varepsilon_{\alert{NS}}^{down}=\frac{2}s\,  p_1\,,$
one obtains the
impact representation for exclusive processes amplitude\footnote{$\kb$ = Eucl. $\leftrightarrow $ $k_\perp$ = Mink.}
\beq
{\cal M} = is\;\int\;\frac{\stmath{d^2\,\kb}}{(2\pi)^2\stmath{\kb^2\,(\rb -\kb)^2}}
\alert{{\Phi}^{\gamma^*(q_1) \to \rho(p^\rho_1)}}(\stmath{\kb,\rb -\kb})\;
\alert{{\Phi}^{\gamma^*(q_2) \to \rho(p^\rho_2)}}(\stmath{-\kb,-\rb +\kb})
\label{impact-rep}
\eq
where 
$\alert{{\Phi}^{\gamma^*(q_1) \to \rho(p^\rho_1)}}$ is the  $\gamma^*_{L,T}(q) g(k_1) \to \rho_{L,T}\, g(k_2)$ impact factor. 

%%%%%%%%%%%%%%%%%%%%%%%%%%%%%%

\subsection{Meson production}

The ''easy'' case (from factorisation point of view) is  $J/\Psi$ production, whose mass provides the required hard scale
\cite{Ryskin:1992uiFrankfurt:1997fjEnberg:2002zyIvanov:2004vd}.
Exclusive vector meson photoproduction at \alert{large $t$} (providing the hard scale)
is another example (which however faces problem with end-point singularities) for which HERA data seems to favor a BFKL picture%~\cite{Enberg:2002zy,
~\cite{Ivanov:2000uqEnberg:2003jwPoludniowski:2003yk}.
Exclusive electroproduction of vector meson  
can also be described~\cite{Goloskokov:2005sdGoloskokov:2006hrGoloskokov:2007nt}
based on improved collinear factorisation for the coupling with the meson DA and collinear factorisation for GPD coupling.

The process $\gamma^{(*)} \gamma^{(*)} \to \rho \, \rho$
is an example of a realistic 
exclusive test of the $\pom$omeron, as  
a subprocess of 
$e^- \, e^+ \to e^- \, e^+ \, \rho_L^0 \, \rho_L^0$ with double lepton tagging, to be made at ILC which should provide the required  very large energy
$(\sqrt{s} \sim 500$ GeV) and 
luminosity ($ \simeq 125\hbox{~fb}^{-1}/\hbox{year}$), with the planned 
detectors designed to cover the \alert{very forward} region, close from the beampipe~\cite{Pire:2005icEnberg:2005eqSegond:2007fjIvanov:2005gnIvanov:2006gtCaporale:2007vs}.

%%%%%%%%%%%%%%%%%%%%%%%%%%%%%%%%%%%%%%%%%%%

\begin{figure}
\psfrag{rh}[cc][cc]{$\hspace{.5cm}\rho$}
\psfrag{gam}[cc][cc]{$\hspace{-1.2cm}\gamma^{(*)}_{T,\,L}$}
\psfrag{PI}[cc][cc]{\scalebox{.7}{$\stmath \!\!\Psi_i$}}
\psfrag{PF}[cc][cc]{\scalebox{.7}{${\aut \!\!\Psi_f}$}}
\psfrag{pi}[cc][cc]{$\! \! p$}
\psfrag{pf}[cc][cc]{$\! \! p$}
\psfrag{sig}[cc][cc]{\scalebox{1.2}{$\alert{\hat{\sigma}}$}}
\psfrag{y1}[cc][cc]{}
\psfrag{fl}[cc][cc]{\scalebox{.85}{\raisebox{-0.35cm}{$\,\updownarrow \!x_{\perp}$}}}
\psfrag{P1}[cc][cc]{}%{\scalebox{.85}{$y_k$}}
\psfrag{PA3}[cc][cc]{}%{\scalebox{.85}{$y_j$}}
%\raisebox{-1cm}{%\vspace{.2cm}
\centerline{\epsfig{file=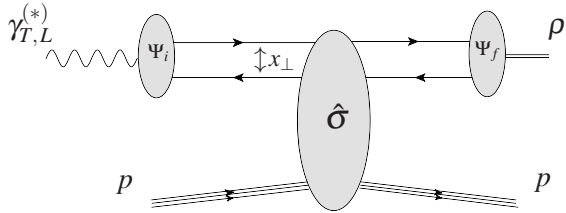,width=6.8cm}}
\caption{Dipole representation for $\gamma^* p \to \rho p$ high energy scattering.}%, where ${\stmath \Psi_i}$ and  ${\aut \Psi_f}$ are the wave functions of projectiles}
\label{Fig:dipole}
\end{figure}

Diffractive vector meson electroproduction have recently been described beyond leading twist, combining collinear factorisation and $k_T-$factorisation.
Based on the $\gamma^*_{L,T} \to \rho_{L,T}$
impact factor
including two- and three-partons contributions, one can describe HERA data on the ratio of the dominant helicity amplitudes~\cite{Anikin:2009hkAnikin:2009bfAnikin:2011sa}.
The dipole representation of high energy scattering~\cite{Mueller:1989stNikolaev:1990ja}
 (Fig.~\ref{Fig:dipole})
is very convenient to implement saturation effects,
through a universal  scattering amplitude $\hat{\sigma}(x_{\perp})$
% \equiv \alert{\hat{\sigma}_{\text{dipole-target}}}$
~\cite{GolecBiernat:1998jsGolecBiernat:1999qd}.
Data for $\rho$ production call for models encoding saturation~\cite{Munier:2001nrKowalski:2006hc}. 
This dipole representation is consistent with the \structure{twist 2} collinear factorisation, and remains valid
beyond leading twist. It seems however that saturation is not enough to describe low $Q^2$ HERA data~\cite{Besse:2012iapheno_saturation}.
The impact parameter dependence provides a probe of the proton shape, in particular through local geometrical scaling \cite{Ferreiro:2002kvMunier:2003bf}.

\subsection{Looking for the Odderon through exclusive processes}

The $\mathbb{O}$dderon, elusive $C-$odd partner of the $\mathbb{P}$omeron, has never been seen in any hard process. One may either  consider exclusive 
processes where the $\mathcal{M}_\mathbb{P}$ amplitude vanishes due to $C$-parity conservation~\cite{Ginzburg:1992miBraunewell:2004pfBzdak:2007cz}
the signal being quadratic in the $\mathcal{M}_\mathbb{O}$ contribution,
 or consider observables sensitive to the \alert{interference} between $\mathcal{M}_\mathbb{P}$ and $\mathcal{M}_\mathbb{O},$ like asymmetries, thus providing observables \alert{linear} in $\mathcal{M}_\mathbb{O}$~\cite{Brodsky:1999mzGinzburg:2002zdGinzburg:2003ciHagler:2002nhHagler:2002sgHagler:2002nfPire:2008xe}.

\section{Conclusion}

Since a decade, there have been much progress in the understanding of hard exclusive processes.  
At medium energies, there is now a conceptual framework starting from first principles, allowing to describe a huge number of processes. At high energy, the impact representation is a powerful tool for describing exclusive processes in diffractive experiments; they are and will be essential for studying QCD in the hard Regge limit ($\pom$omeron, $\odd$dderon, saturation...). 
Still, some problems remain:
proofs of factorisation have been obtained only for very few processes
(ex.: $\gamma^* \, p \to \gamma \, p\,$, $\gamma^*_L \, p \to \rho_L \, p$). For some other processes, it is highly plausible, but\linebreak  not fully demonstrated, like those involving GDAs and TDAs. Some processes explicitly show sign of breaking of factorisation
(ex.:  $\gamma^*_T p \to \rho_T p$  at leading order).
The effect of QCD evolution, the NLO corrections and the choice of renormalization/factorisation scale~\cite{Brodsky:1982gcAnikin:2004jbMoutarde:2013qs}, as well as power corrections will be very relevant to interpret and describe the forecoming data. 
The AdS/QCD correspondence may provide insight for modeling the involved non-perturbative correlators~\cite{Brodsky:2006uqaGao:2009seMarquet:2010sf}.

\end{document}